\documentclass[conference]{IEEEtran}
\IEEEoverridecommandlockouts
\usepackage{graphicx}
\usepackage{tabularx}
\usepackage[justification=centering]{caption}
\usepackage{cite}
\usepackage{gensymb}
\usepackage{graphicx}
\usepackage{tabularx}
\usepackage[justification=centering]{caption}
\usepackage{cite}
\usepackage{latexsym}
\usepackage{amsmath}
\usepackage{algorithm}
\usepackage[noend]{algpseudocode}
\usepackage{booktabs}
\usepackage{multirow}
\usepackage{siunitx}
\usepackage{pdfpages}
\usepackage{array,booktabs}
\usepackage{amsmath}
\usepackage[utf8]{inputenc}
\usepackage{mathtools}
\usepackage{amssymb}
\usepackage{pdfpages}
\usepackage{booktabs}
\usepackage{afterpage}
\usepackage{float}
\usepackage{bbding}
\usepackage{pifont}
\usepackage{wasysym}
\usepackage{subfigure}
\usepackage{amsfonts}
\usepackage{siunitx}
\usepackage{xcolor}
\usepackage{mathtools, cases}
\usepackage{float}
\usepackage{stfloats}
\usepackage{caption}
\newcolumntype{P}[1]{>{\centering\arraybackslash}p{#1}}
\newcolumntype{M}[1]{>{\centering\arraybackslash}m{#1}}
\DeclareCaptionLabelFormat{continued}{#1~#2}
\renewcommand{\baselinestretch}{0.975}

\usepackage{atbegshi}
\AtBeginDocument{\AtBeginShipoutNext{\AtBeginShipoutDiscard}}

\begin{document}


\title{LTE and Wi-Fi Coexistence in Unlicensed Spectrum with Application to Smart Grid: A Review }


\author{Yemeserach Mekonnen, Muhammad Haque, Imtiaz Parvez, Amir Moghadasi, Arif Sarwat\\
Department of Electrical and Computer Engineering, Florida International University, Miami, FL\\
Email: \tt\{ymeko001,mhaqu007,iparv001,amogh004,asarwat\}@fiu.edu}

\thanks{This research was supported in part by the U.S. National Science Foundation under the grant RIPS-1441223 and CAREER-0952977.}
\maketitle

\begin{abstract}
Long Term Evolution (LTE) is expanding its utilization in unlicensed band by deploying LTE Unlicensed (LTE-U) and Licensed Assisted Access LTE (LTE-LAA) technology. Smart Grid can take the advantages of unlicensed bands for achieving two-way communication between smart meters and utility data centers by using LTE-U/LTE-LAA.  However, both schemes must co-exist with the incumbent Wi-Fi system. In this paper, several co-existence schemes of Wi-Fi and LTE technology is comprehensively reviewed. The challenges of deploying LTE and Wi-Fi in the same band are clearly addressed based on the papers reviewed.  Solution procedures and techniques to resolve the challenging issues are discussed in a short manner. The performance of various network architectures such as listen- before-talk (LBT) based LTE, carrier sense multiple access with collision avoidance (CSMA/CA) based Wi-Fi is briefly compared. Finally, an attempt is made to implement these proposed LTE-Wi-Fi models in smart grid technology.

\end{abstract}

\begin {IEEEkeywords}
LTE, Wi-Fi, Coexistence mechanism, LTE-LAA, LTE-U, LBT, Smart grid
\end{IEEEkeywords}


\section{Introduction}
The surge in smart phones, tablets, mobile APs, wearable and Internet-of-Things (IoT) has exponentially increased mobile data usage and wireless communication resulting in huge explosion of traffic demand on the licensed frequency bands. Similarly, the smart grid network has added more traffic to the existing channels which necessitate utilizing more frequency bands to increase  capacity of the operators to fulfill the drastic demand of the traffic. Having large amount of radio resources, the unlicensed spectrum are recently treated as an excellent supplementary frequency bands to augment the throughput of wireless communications [1]. LTE is generally divided into LTE-U and LTE-LAA when used in unlicensed spectrum [2]. LTE-U was the early deployment which has simple mechanism and doesn’t require alteration to existing LTE air interface protocol. It employs LTE Release 10-12 aggregation protocol which doesn’t require the Listen Before Talk (LBT) algorithm [3]. It is only applicable in the US, South Korea, India and China market rather than in Europe and Japan which are LBT regulated markets. LAA is ratified by the 3rd Generation Partnership (3GPP) as Release 13 which only aims on single global framework [4, 5]. 

\begin{figure}[ht!]
   \centering
   \includegraphics[scale=.45]{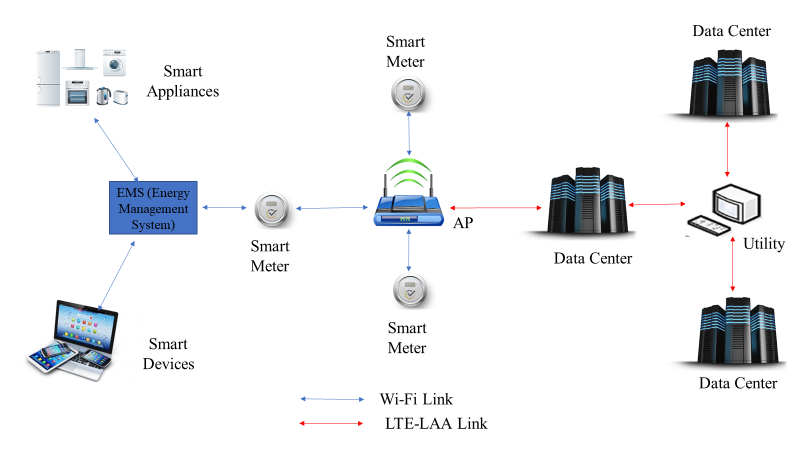}
    \caption{Wi-Fi-LTE based Smart Grid Communication}
    \label{fig:smart_grid}
\end{figure}
\setlength{\textfloatsep}{0.3\baselineskip plus 0.2\baselineskip minus 0.5\baselineskip}

In addition to meeting the LBT regulation, LAA meets the channel occupancy bandwidth and power spectral density. LAA enables LTE to transmit data via both licensed and unlicensed bands simultaneously. However, deploying LTE in unlicensed bands can cause Wi-Fi, the incumbent systems in unlicensed spectrum, to be severely interfered. For this reason, coexistence schemes are needed to avoid remarkable efficiency deterioration by neighboring LTE and Wi-Fi channels.

\begin{table*}[hb] 
\caption{A Short Comparison between LTE-U and LTE-LAA} 
\centering 
\begin{tabular}{>{\centering\arraybackslash}p{2cm} >{\centering\arraybackslash}p{6cm} >{\centering\arraybackslash}p{6cm}}
\hline 
Index &LTE-U & LTE-LAA\\ 
\hline 
$Operation [32]$ & Uses both UL and DL frequency in licensed spectrum while only uses SDL with carrier aggregation in unlicensed spectrum & Suitable to be employed in both licensed and unlicensed bands by increasing bandwidth and bitrate with carrier aggregation. \\  
\hline
$Differences$\\$[32-34]$\
\vline

& 
\begin{itemize}
\vspace*{-\baselineskip}
\item Based on Release 10, 11 and 12
\item Does not use LBT algorithm
\item Aggregate throughput of two co-existing LTE-U networks is larger than two coexisting Wi-Fi networks
\item Operate in both 900 MHz and 5 GHz band
\item Shows less coverage and capacity while coexists with Wi-Fi as compared with LAA
\item Probability of Noise and interference is higher 
\item Indoor throughput is less than outdoor throughput 
\end{itemize} &
\begin{itemize}
\vspace*{-\baselineskip}
\item Based on Release 13
\item Use LBT algorithm
\item Aggregate throughput of two coexisting LAA networks is larger than two coexisting Wi-Fi networks. 
\item Special operating band is 5 GHz.
\item Shows better coverage and capacity while coexists with Wi-Fi
\item Probability of Noise and interference is comparatively low
\item Indoor and outdoor throughput both show better improvement
\end{itemize}
\\
\hline
$ Coexistence$\\ $Mechanism $\ & 
\begin{itemize}
\vspace*{-\baselineskip}
\item Channel selection
\item CSAT (Carrier Sense Adaptive Transmission)
\item Opportunistic Supplemental Downlink
\end{itemize}
 & 
\begin{itemize}
\vspace*{-\baselineskip}
\item FBE based LBT
\item LBE based LBT
\end{itemize}
\\  
\hline  
\end{tabular} 
\label{tab:var} 
\vspace*{-4mm}
\end{table*}

There are different coexistence systems proposed in [6] through experimental analysis where the authors confirm that LTE significantly impacts Wi-Fi performance in different environment scenarios. The Downlink (DL) performance of coexisting LAA and Wi-Fi is analyzed in [7] by using 3 Markov chain models to compute the throughput and delay of Wi-Fi networks and Wi-Fi-LAA networks. In [8-10], LBT based MAC protocol is modeled for the LTE-LAA system so that the Wi-Fi performance in the coexistence system can be regulated by tuning the transmission time of LTE-LAA only. The studies in [11] proposed a stochastic geometry model for co-existing Wi-Fi and LTE-LAA to measure some of their key performance metrics including the Medium Access Probability (MAP), the signal-to-interference-plus-noise ratio (SINR) distribution, the Density of Successful Transmissions (DST). In [12], Multiple Signal Classification (MUSIC) and Direction of Arrival (DOA) estimation is combined with null steering techniques to eleminate interference between LTE-LAA and Wi-Fi datatransfer. The authors in [13-14] analytically derived an optimistic radio accessibility for LAA to model a dynamic switching between scheduling-based and random access architecture. The performance of an LBE-based LTE and Carrier Sense Multiple Access/Collision Avoidance (CSMA/CA) under similar conditions are analyzed in [15-19] and it’s found that the efficiency of both schemes is significantly dependent on the number of nodes.
However, none of the above papers talked about the implementation of Wi-Fi-LTE coexistence in smart grid. The authors in [20-21] analyzed and reviewed the application of Wi-Fi-LTE based systems in smart grid technology and proposed some solutions to overcome the drawback issues coming out from Wi-Fi-LTE co-existence. In this study, the co-existence challenge of LTE-LAA and Wi-Fi in the unlicensed frequency is investigated.

This paper will serve as a review work on techniques that have been proposed previously on the coexistence issue of LTE-LAA and Wi-Fi networks. Solution schemes to resolve the co-existence challenge modeled in different papers are summarized. It will further attempts to describe how these co-existence schemes will be implemented on smart grid to facilitate communication. Encounters of deploying LTE-Wi-Fi in smart grid data transfer and their solutions are discussed.

The rest of this paper is organized as follows. In Section~II, a general overview of Wi-Fi and LTE is presented. Section~III describes the major challenges of coexistence of Wi-Fi and LTE. Two different coexistence mechanism are presented in Section~IV. The coexistence techniques are reviewed from literature in Section~V. Finally, concluding remarks are presented in Section~VI.


\section{Short Overview of Wi-Fi and LTE}

	\subsection{Wi-Fi Technology}
	    Wi-Fi is a WLAN (Wireless Local Area Network) technology, which is vastly used to link up different types of wireless devices like smart phones, laptops, tablets, and other smart devices to each other with the help of Internet via a Wi-Fi Access Point (AP). Advantages like easy set-up, high data rate, short round trip delay and low cost, make Wi-Fi the most popular WLAN technology. However, a certain number of devices are permitted to be connected to a specific Wi-Fi network due to its limited number of channel accessibility to avoid any interference [13].
 
Wi-Fi technology uses Medium Access Control(MAC) layer mechanism to be deployed in unlicensed spectrum. MAC is the protocol layer of Wi-Fi which controls and maintains different Wi-Fi devices in the Wi-Fi networks by coordinating their access to a shared channel. Several MAC Wi-Fi mechanisms are proposed in the industry, however, the most popular one is Distributed Coordination Function (DCF) [15]. It’s based on CSMA/CA technology in which the Wi-Fi device first senses the channel before making a transmission, and completes a Clear Channel Assessment (CCA) mechanism. In this mechanism, a channel is checked if it's idle or free for a certain period called DIFS (Distributed Inter- Frame Space), the device starts to transmit. During the transmission, if a collision happens, the Wi-Fi device enters back-off mode and waits for a random time before trying again.
	\subsection{LTE Technology}
	 LTE is a mobile communication architecture which was initially designed for the communication of wireless devices and data terminals in the licensed bands.Having high spectral proficiency, improved peak data rates, short round trip time and frequency flexibility, LTE is widely called as 4G in the industry [15]. This standard was designed by 3rd Generation Partnership Project (3Gpp).

    The Physical (PHY) layer of an LTE has Down Link (DL) and Up Link (UL) features. The PHY layer necessitates high peak transmission rates, several channel bandwidths and spectral efficiency. The access method LTE uses is based on Orthogonal Frequency Division Multiple Access (OFDMA) with a mixing of larger bandwidth modulation and spatial multiplexing in the DL [24]. That’s why, the highest peak data rate of LTE Rel. 8 in the UL is 75 Mbps, while in the DL it can support up to 300 Mbps. However, LTE Release 8 could not meet the expected technical criteria of 4G wireless service which results in LTE Rel. 10, a more advanced technology popularly called as LTE-A (LTE Advanced). LTE-A is the first release to include data offloading from LTE to the unlicensed spectrum [25]. 
     Presenting LTE in the unlicensed band enables service providers to improve their existing services and keep up with the growing demand. So, the industry came up with several types schemes for deploying LTE in the unlicensed spectrum. Among them, LTE-U and LTE-LAA are discussed in this paper. A short comparison of LTE-U and LTE-LAA is summarized in Table~I.

\section{Coexistence Challenges in smart grid}

     The current smart grid network consists of smart meters, numerous monitoring systems and sensors. The Advanced Metering Infrastructure (AMI) is one of the building blocks of the smart grid consisting of numerous smart meter networks from the distribution end. AMI's prominent purpose is to act as data manager facilitating flow of information and enabling bi-directional communication in the network. The networking mechanism of  AMI are divided into Home Area Network (HAN) and Neighborhood Area Networks (NAN) [22]. HAN uses Wi-Fi for data transferring between smart home appliances and Wi-Fi AP while NAN uses LTE for the communication between AP and utilities. Fig.1 shows a Wi-Fi-LTE combination in smart grid communication based on the coexistence model designed in [20]. Smaller data rate and lower interference tendency make Wi-Fi to be used in HAN. Having superior quality than LTE-U in unlicensed band, LTE-LAA has two important application in smart grids: automated metering and controlling the distribution system [26]. LTE-U/LTE-LAA networks can Utilize the large bandwidth of unlicensed spectrum to process a sophisticated communication for Automated Demand Response (ADR), AMI and outage restoration.
     
Challenges of Wi-Fi-LTE co-existence are very prominent in smart grids. Firstly, the Wi-Fi network are more susceptible to electromagnetic interference due to high electrical voltage equipment and from other wireless equipment which degrades performance. Secondly, despite the advent of smart meters, wireless equipment is not readily available giving rise to compatibility issues [20]. However, Wi-Fi have gathered advanced acknowledgment protocols, error correction algorithms and data  buffering  to increase its reliability in smart grid communications.
The main feature of LTE is to expedite Down Link channel and VOIP (voice over Internet Protocol) traffic profiles, however, having different characteristics, smart meter traffic profile can pose challenges for LTE communications. The authors in [19] addressed some frequent challenges that LTE faces in smart grid communications such as Physical Downlink Control Channel (PDCCH), Physical Downlink Shared Channel (PDSCH) and Physical Uplink Shared Channel (PUSCH). Additionally, a large number of LTE devices can cause enormous competition for accessing into the channel and decrease the capability of Evolved Node-B (eNB) to gather correct information about the user equipment (EU).
	

\section{Coexistence mechanism}
In this section, two LTE-U coexistence scenarios are discussed. The first case discusses the coexistence between Wi-Fi and LTE, and the second coexistence between LTE of different operators. Unlike Wi-Fi, LTE devices do not perform carrier sensing before transmission. This is a primary reason why coexistence issues arise and why different types of coexistence mechanisms had to be developed for LTE to avoid the interference with Wi-Fi.

\subsection{LTE vs Wi-Fi}
    One of the main differences of these two systems are radio frame structure and transmission scheduling [23].  LTE-U and Wi-Fi also use different MAC/PHY designs and are most times operated by different operators. Both systems use different channel access mechanisms leading to the main difference in the coexistence of these systems [24]. LTE are used in licensed band, and has centralized controller for DL/UL having high spectrum efficiency and transmissions.  On the other hand, Wi-Fi systems don’t require centralized control, and use DCF based on random access. 
When it comes to channel usage, LTE transmits for neighboring frames and the channels are on. However, Wi-Fi systems send packets only when channel is unoccupied [1]. Therefore, when LTE and Wi-Fi systems use the same unlicensed band, the performance of Wi-Fi will be greatly impacted.

Let us consider, a collocated LTE and Wi-Fi network scenario  where LTE and Wi-Fi coexists in the same unlicensed band. We consider that the sets of Wi-Fi APs, LTE-U BS, Wi-Fi STAs and LTE-U UE are given by $S_w$, $S_l$, $U_w^i$ and $U_l^j$ respectively. The transmission power of LTE BS $x$, LTE BS $a$, Wi-Fi AP $i$ and Wi-Fi STA $b$ are $p_r^x$, $p_r^a$, $p_r^i$ and $p_r^b$.

The channel gain values from LTE BS $x$ to LTE UE $j$, from  LTE BS $a$ $(a \neq x)$ to LTE UE $j$, from WiFi AP $i$ to LTE UE $j$ and Wi-Fi STA $b $ to Wi-Fi $j$ are $h_{j,r}^x$, $h_{j,r}^a$, $h_{j,r}^i$ and $h_{j,r}^b$ respectively.

The signal-to-noise (SINR) of LTE UE $j$ on resource block $r$ during receiving data from LTE BS $x$ \cite{I29} is

\begin{equation}
\mathrm{SINR}_{j,r}^x=\frac{h_{j,r}^x p_r^x}{\sum{h_{j,r}^a p_r^a}+\sum {h_{j,r}^i p_r^i}+\sum {h_{j,r}^b p_r^b}+\sigma^2}
\end{equation}

The number of successful transmission bit $N_{\mathrm{B}}$ from the LTE BS $x$ to LTE UE $j$ is

\begin{equation}
N_B^x= \mathrm{BT} \sum \log_2 (1+\mathrm{SINR}_{j,r}^x)
\end{equation}

where B is the bandwidth and 
T is the transmission time such that T=$\sum r$.\\

The down link (DL) capacity of LTE UE $x$

\begin{equation}
C^x=\frac{N_B^x}{T_{Tx}+T_{wait}}
\end{equation} 

Where $T_{Tx}$ and $T_{wait}$ are the transmission and wait time, respectively. The wait time may be due to the listen before talk (LBT) \cite{c41} or duty cycled transmission period \cite{c14}. \\

	\subsection{LTE-U vs LTE-U}
    This case highlights when LTE-Us from different operators coexist in the same unlicensed spectrum [4]. In markets, not regulated by LBT, neighboring LTE-U eNB utilize the same unlicensed carrier at the same time causing a great co-channel interference especially when the different operators are not well coordinated in the same deployment area [15,25]. LBT requires the application of a CCA check before using the channel. It is a fair way of sharing spectrum in unlicensed bands where other systems operate simultaneously with LTE [25]. In addition, when multiple LTE-U nodes try to identify a clear unlicensed spectrum simultaneously, interference problem among the nodes arises. Therefore, a proper access mechanism of LBT must be implemented to reduce interference and transmission reliability. In LBT regulated markets, a fair competition principle such as online auction mechanism can be an optional function to improve performance. This assures that LTE-U nodes can be coordinated in a fair manner to share the unlicensed spectrum efficiently.

\section{Techniques for Coexistence}

Huge difference between LTE and Wi-Fi pose great challenges in the design of an effective coexistence mechanism. The above factors should be carefully studied to design a fair and efficient coexistence mechanism for LTE and Wi-Fi networks in unlicensed band. 
	\subsection{Coexistence without LBT}
    In countries where there is no regulatory requirement for LBT, careful designed coexistence algorithm will guarantee a fair coexistence. Using the Release 10/11 LTE PHY/MAC standards, three mechanisms can be implemented to safeguard that LTE’s coexistence in unlicensed band with Wi-Fi. Channel Selection permits smalls cells to choose the clearest channel based on Wi-Fi and LTE analysis [23]. If clear channel is found, LTE–U will occupy with full duty circle for Secondary DL (SDL) transmission. If no clean channel is available, Carrier-Sensing-Adaptive Transmission (CSAT) is used to share the channel with Wi-Fi [17]. Depending on traffic demand, SDL carrier can be opportunistically retrieved. If there is low load, SDL carrier should be turned off and for higher load, channel selection should be executed again.

    \begin{table*}[!tp] 
    \caption{Summary of Coexistence Mechanisms and Approach} 
    \centering 
    \begin{tabular}{cll}
    \hline
    \begin{tabular}[c]{@{}c@{}}Coexistence \\ Mechanism\end{tabular}                    & \multicolumn{1}{c}{Techniques}                                           & \multicolumn{1}{c}{Approach}                                                                                                                                                                                                                                                                                                                                                                      \\ \hline
    \multirow{3}{*}{\begin{tabular}[c]{@{}c@{}}Without LBT\\ {[}2{]}{[5}{]}\end{tabular}} & \begin{tabular}[c]{@{}l@{}}Channel Selection\\ {[}21{]}{[25}{]}\end{tabular} & \begin{tabular}[c]{@{}l@{}}Scans fro the cleanest channel in the unlicensed band for SDL transmission\\ Interference level is measured by energy detection\end{tabular}                                                                                                                                                                                                                           \\ \cline{2-3} 
                                                                                        & CSAT                                                                     & \begin{tabular}[c]{@{}l@{}}Depending on medium activities, small cells gates off LTE transmission \\ Provide coexistence across different technologies in TDM mode\\ Utilizes adaptive duty cycle LTE-U and Wi-Fi ON/OFF state\\ Similar to CSMA algorithm but has longer latency\end{tabular}                                                                                                    \\ \cline{2-3} 
                                                                                        & Opportunistic SDL {[}23{]}                                               & \begin{tabular}[c]{@{}l@{}}Dependent on traffic and transmission load demand \\ During higher traffic  and active users exist, SDL will be turned ON\\ When primary carrier can mange traffic and no active users exist, SDL is OFF\end{tabular}                                                                                                                                                  \\ \hline
    \multirow{2}{*}{\begin{tabular}[c]{@{}c@{}}With LBT \\ {[}18{][}21{]}\end{tabular}}   & FBE {[}17{]}                                                             & \begin{tabular}[c]{@{}l@{}}Equipment where transmit/receive structure has fixed time frame using CCA algorithm\\ It is implemented every fixed frame period \\ The pros are simple design,  and less standardization compared to LBE\end{tabular}                                                                                                                                                 \\ \cline{2-3} 
                                                                                        & LBE                                                                      & \begin{tabular}[c]{@{}l@{}}It is demand driven and apply CCA using energy detect for channel observation time 20. \\ If clear, transmit with channel occupancy time, if not it executes extended CCA\\ Unlike FBE, can easily access channel once clear and has superior efficiency and minor delay.\\ Cons are collision rate and channel access degrades for large number of users.\end{tabular} \\ \hline
    \end{tabular}
    \end{table*}

\subsubsection{Channel Selection}    
In this mechanism, LTE-U small cells will scan the unlicensed band to search for the cleanest unused channels for the SDL carrier transmission. Given that there is an unused channel, the interference is avoided between the cells and its nearby Wi-Fi devices and other LTE-U small cells. Operating channel is monitored on an on-going basis by Channel Selection Algorithm.  Measurements are usually completed at both the beginning power-up stage and later periodically at SDL operation stage. This period is usually at 10s of seconds. When interference is detected in the operating channel, LTE-U will attempt to switch to another clear channel with less interference based on LTE Release 10/11 procedures[1,4,18]. 
The interference level in a channel is measured by energy detection where initially the quantity of interference sources and types are unknown. LTE and Wi-Fi measurements are engaged to augment interference detection. 

\subsubsection{Carrier-Sensing-Adaptive Transmission (CSAT)} 

When no clean channel is available, LTE-U will be able to share the channel by implementing adaptive duty cycle or CSAT algorithm. The aim of the CSAT algorithm is to afford coexistence across different technologies in a Time Division Multiplexing (TDM) mode[10]. In general, the coexistence methods in unlicensed band are by using LBT or CSMA for Wi-Fi, which uses contention based access. For CSMA or LBT, the medium should be sensed and accessed if it is clear in order to implement TDM for coexistence. LTE-U radio continue measuring occupancy on a channel and decide how many frames to transmit or how many to stay quiet which is known as duty cycle. Duty cycle facilitate the interaction when LTE-U is ON and Wi-Fi is in OFF state. The LTE-U, which is on secondary cell is occasionally activated and de-activated using LTE MAC control elements.

\subsubsection{Opportunistic Supplementary Downlink (SDL)}

This mechanism is dependent on traffic and load demand. If the DL traffic of the small cell exceeds certain threshold and there exist active users within the unlicensed band spectrum, the SDL carrier can be turned on for offloading. On the other hand, when the primary carrier can easily manage the traffic demand and there is no user within the unlicensed band coverage the SDL is turned off. Opportunistic SDL decreases the interference from continuous RS transmission from LTE-U in unlicensed channel subsequently leading in noise reduction in and around a shared channel [2]. 
    
\subsection{Coexistence Based on LBT Mechanism}
For Europe, Japan and India markets that requires a regulation in the unlicensed spectrum require a more robust equipment to periodically check for presence of other occupants in the channel(listen ) before transmitting (talk) in millisecond scale. Two LBT mechanisms are employed in LTE-LAA mandated by European Telecommunications Standards Institute (ETSI). One is Frame based Equipment (FBE) and Load based Equipment (LBE) [11,12].

\subsubsection{FBE-Based LBT Mechanism}

In this mechanism, the equipment has a fixed frame period, where CCA is executed. When the current operating channel is dimmed to be clear, the equipment immediately can transmit for duration equivalent to the channel occupancy time [21]. Similarly, if the operating channel is busy, the equipment cannot transmit on the channel for the next fixed frame period. The channel occupancy time requirement is minimum 1ms and maximum 10ms and idle period accounting 5 of channel occupancy time. FBE-based LBT is simple for the design of reservation signal and requires less standardization.

\subsubsection{LBE-Based LBT Mechanism}

In LBE, the equipment is required to define whether the channel is clear or not. Unlike FBE, LBE is demand-driven and not dependent on fixed time frame. In the case where the equipment discovers a clear operating channel, it will instantly transmit. If not, an Extended CCA (ECCA) is implemented, where the channel is observed for a period of random factor N multiplied by the CCA time slot [16,18]. N is the quantity of clear slots so that a total idle period should be observed before transmission. Its value is chosen randomly from 1 to q, where q has a value from 4 to 32. When a CCA slot is idle, the counter will be cut by one. The equipment can transmit even if the counter reaches zero. In addition, the maximum channel occupancy time is calculated by (13/32) *q ms. Therefore, the maximum channel occupancy time is 13 ms when q equals to 32 which is the best coexistence parameter.

The overview of LBT and non-LBT techniques are summarized in Table~II.


\section{Conclusion}

It is observed that Wi-Fi performance is degraded while LTE is slightly impacted in coexistence scenario. When in coexistence, Wi-Fi networks are more likely to be blocked by LTE transmission and other wireless equipment. The coexistence mechanism discussed in this paper ensures a fair and friendly coexistence in the unlicensed spectrum. Special detail and explanation is given to two different coexistence mechanisms for functioning LTE in unlicensed bands in regions with and without LBT requirements. For LBT, the design of an effective coexistence mechanism is critical to the dense deployment of the advanced metering infrastructure in smart grid networks where the LTE-Wi-Fi networks can be utilized. For coexistence with out LBT market, channel selection, CSAT and SDL ensures Wi-Fi-LTE systems friendly functionality. For markets with LBT regulatory requirements, FBE and LBE mechanisms are discussed. LBE based LBT has superior efficiency than FBE based LBT in terms of resource page and minor delay.

\end{document}